\documentclass[10pt]{iopart}
\usepackage{epsf,graphicx,amssymb}
\newcommand{\be}{\begin{equation}}
\newcommand{\ee}{\end{equation}}

\begin{document}
\title[Correlated multiplexity and connectivity of multiplex random networks]{Correlated multiplexity and connectivity of multiplex random networks}
\author{Kyu-Min~Lee, Jung Yeol Kim, Won-kuk~Cho, K.-I.~Goh,$^*$ and I.-M.~Kim}
\address{Department of Physics and Institute of Basic Science, Korea University, Seoul 136-713, Korea}
\ead{$^*$kgoh@korea.ac.kr}
\date{\today}
\begin{abstract}
Nodes in a complex networked system often engage in more than one type of interactions among them;
they form a {\it multiplex} network with multiple types of links. 
In real-world complex systems, a node's degree for one type 
of links and that for the other are not randomly distributed but correlated,
which we term {\it correlated multiplexity}.
In this paper we study a simple model of multiplex random networks
and demonstrate that the correlated multiplexity can drastically affect the properties of giant component in the network.
Specifically, when the degrees of a node for different interactions in
a duplex Erd\H{o}s-R\'enyi network are maximally correlated, the network 
contains the giant component for any nonzero link densities.
On the contrary, when the degrees of a node are maximally anti-correlated,
the emergence of giant component is significantly delayed, yet
the entire network becomes connected into a single component at a finite link density.
We also discuss the mixing patterns and the cases with imperfect correlated
multiplexity.
\end{abstract}
\maketitle

\section{Introduction}
In the last decade, network has proved to be a useful framework
to model structural complexity of complex systems \cite{small,scale}.
By abstracting a complex system into nodes (constituents)
and links (interactions between them), the resulting graph 
could be efficiently treated analytically and numerically,
through which a large body of new physics of complex systems 
has been acquired \cite{caldarelli-book,newman-book,havlin-book}.
Most studies until recently have focused on the properties
of isolated, single networks where nodes interact with a single type of links.
In most, if not all, real-world complex systems, however, nodes
in the system can engage in more than one type of interactions or links;
People in a society interact via their friendship, family
relationship, and/or more formal work-related links, {\it etc.}
Countries in the global economic system interact via various financial
and political channels ranging from commodity trade to political alliance.
Even proteins in a cell participate in multiple layers of interactions
and regulations, from transcriptional regulations and metabolic synthesis 
to signaling.
Therefore, a more complete description of complex systems
would be the {\it multiplex} network \cite{wasserman} with more than
one types of links connecting nodes in the network.
Multiplexity can also have impact on network dynamics;
many dynamic processes occurring in complex network systems
such as behavioral cascade in social networks \cite{charlie} or dynamics of
systemic risk in the global economic system \cite{kmlee} should
be properly understood from the perspective of multiplex network dynamics.
Since its introduction in the social network literature \cite{wasserman},
however, only a handful of earlier related studies have existed
in the physics literature, notably Refs.~\cite{soderberg,dhkim,kurant},
and the understanding of generic effects of multiplexity remained lacking.

More recently, related concepts such as interacting networks \cite{leicht}
and interdependent networks \cite{buldyrev} have been
introduced and studied. 
Leicht and D'Souza \cite{leicht} studied what they called 
interacting networks, in which two networks are coupled via inter-network edges,
and developed a generating function formalism to study their percolation 
properties.
Buldyrev {\it et al.}~\cite{buldyrev} studied the interdependent networks, 
in which mutual connectivity in two network layers plays an important role, 
and found that catastrophic cascades of failure can occur 
due to the interdependency. 
Although the specific contexts are different in these studies, 
if one regards each type of links in a multiplex network to constitute
a network layer, and the multiplex network as multilayer network, the multiplex
networks and the interacting or interdependent networks may be described
by a similar framework at the mathematical level.
In this sense, these studies have provided a pioneering insight 
relevant to multiplex networks that there can be nontrivial effects 
of having more than one type of links, or channel of interactions, 
in networks \cite{parshani-pnas,brummitt}.  

In Refs.~\cite{leicht,buldyrev}, network layers were coupled in 
an uncorrelated way, in the sense that
the connections or pairings between nodes in different layers are 
taken to be random. 
In real-world complex systems, however, 
nonrandom structure in network multiplexity can be significant.
For example, a person with many links in the friendship layer
is likely to also have many links in another social network layer,
being a friendly person. Such a nonrandom, or correlated multiplexity
has recently been observed in the large-scale social network analysis of
online game community \cite{szell}, in the world trade system \cite{garlaschelli},
and in transportation network systems \cite{inter-similarity,he}, and its impact on network robustness has been studied \cite{inter-similarity,buldyrev2}.
In this paper, our main goal is to understand the generic role
of correlated multiplexity in multiplex system's connectivity structure.

\section{Model and formalism}
\begin{figure}
\includegraphics[width=.9\linewidth]{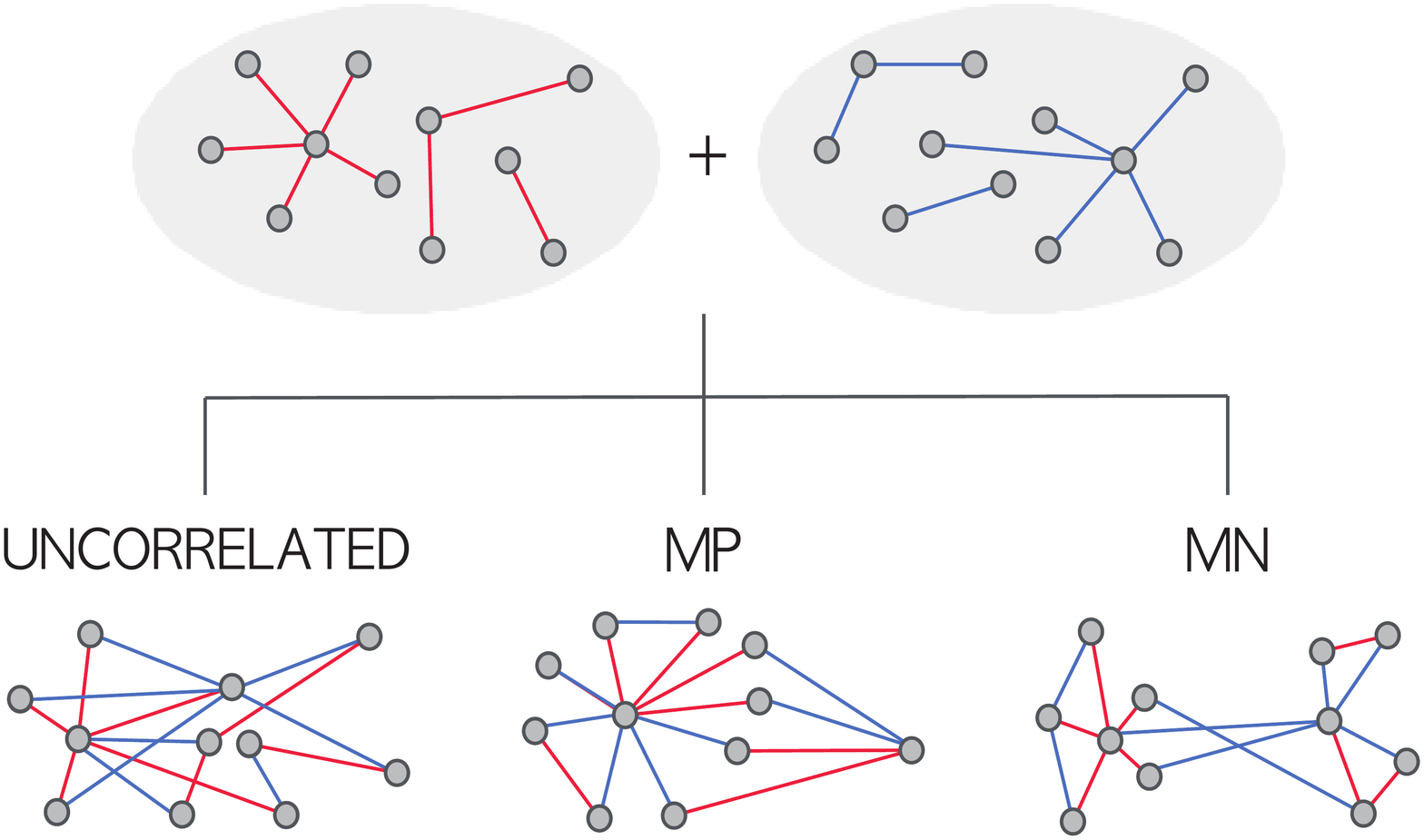}
\caption{Schematic illustration of multiplex ER networks
with three kinds of multiplex couplings discussed in the text. MP (MN) stands for maximally-positive (maximally-negative) correlated multiplexity.}
\end{figure}

To study how the network connectivity is affected
by correlated multiplexity, we consider the following model of 
multiplex networks with two layers, or {\it duplex} networks.
The network has $N$ nodes connected by two kinds of links,
modeling, for example, individuals participating in 
two different social interaction channels.
We refer the subnetwork formed by each kind of links to as the network layer.
Each network layer $l$ ($l=1,2$) is specified by the intralayer
degree distribution $\pi^{(l)}(k_l)$, where degree $k_l$ is the
number of links within the specific layer $l$ of a node.
The complete multiplex network can be specified by the joint distribution 
$\Pi(k_1,k_2)$ or the conditional degree distribution $\Pi(k_2|k_1)$. 
Generalization into $l>2$ layers is straightforward.

The total degree of a node in the multiplex network
is given by $k=k_1+k_2-k_{o}$, where $k_o$ denotes the number of
overlapped links in the two layers, which can be neglected in the $N\to\infty$ limit for random, sparse networks with largest degree of at most order ${\cal O}(\sqrt{N})$ \cite{nakata}.
One can obtain the total degree distribution $P(k)$ from
the joint degree distribution or the conditional degree distribution as 
$P(k)=\sum_{k_1,k_2}\Pi(k_1,k_2)\delta_{k,k_1+k_2}=\sum_{k_1}\Pi(k-k_1|k_1)\pi^{(1)}(k_1)$,
where $\delta$ denotes Kronecker delta symbol.
From $P(k)$, one can follow the standard generating function 
technique \cite{newman} to study the network structure:
The generating function $g_0(x)$ of
the degree distribution $P(k)$ of the mutiplex network 
can be written as
\be
g_0(x)=\sum_{k=0}^{\infty}P(k)x^{k}=\sum_{k_1,k_2}\Pi(k_1,k_2)x^{k_1+k_2}~.
\ee
Emergence of the giant component spanning a finite fraction of the network 
signals the establishment of connectivity.
Size of the giant component $S$ is obtained via 
\be S=1-g_0(u)=1-\sum_{k=0}^{\infty}P(k)u^k, \ee
where $u$ is the smallest root of the equation 
$x=g'_0(x)/g'_0(1)\equiv g_1(x)$, that is,
\be u=\frac{1}{\langle k\rangle}\sum_{k=1}^{\infty}kP(k)u^{k-1}. \ee
Mean size of the component to which a randomly chosen vertex belongs, 
$\langle s\rangle$, plays the role of susceptibility and is given by
\be \langle s\rangle=1+\frac{g'_0(1)u^2}{g_0(u)[1-g'_1(u)]}~.\ee
The condition for existence of the giant component (that is, $S>0$)
is given by the existence of a nontrivial solution $u<1$, leading
to the so-called Molloy-Reed criterion \cite{newman,molloy,cohen}, 
\be \sum_kk(k-2)P(k)=\langle k^2\rangle-2\langle k\rangle>0~. \ee 
It is worthwhile to note the related recent generalizations of the generating function method for interacting \cite{leicht} and interdependent networks \cite{son}.

For a multiplex network system, the total degree distribution $P(k)$ 
is determined from the joint degree distribution
$\Pi(k_1, k_2)$, which depends on the pattern of correlated multiplexity.
Therefore, the presence of correlated multiplexity
affects the multiplex system's connectivity (figure 1).
In the following, we specifically consider 
duplex networks of two Erd\H{o}s-R\'enyi (ER) layers \cite{er} 
and three limiting cases of correlated multiplexity.
Using analytical treatment with mean-field-like approximation
as well as extensive numerical simulations,
we present how the correlated multiplexity can affect the emergence of the giant component in the multiplex system.
Similar procedures can also be applied to multiplex scale-free network 
models \cite{static}.

\section{Degree distributions}

\begin{figure}
\includegraphics[width=.9\linewidth]{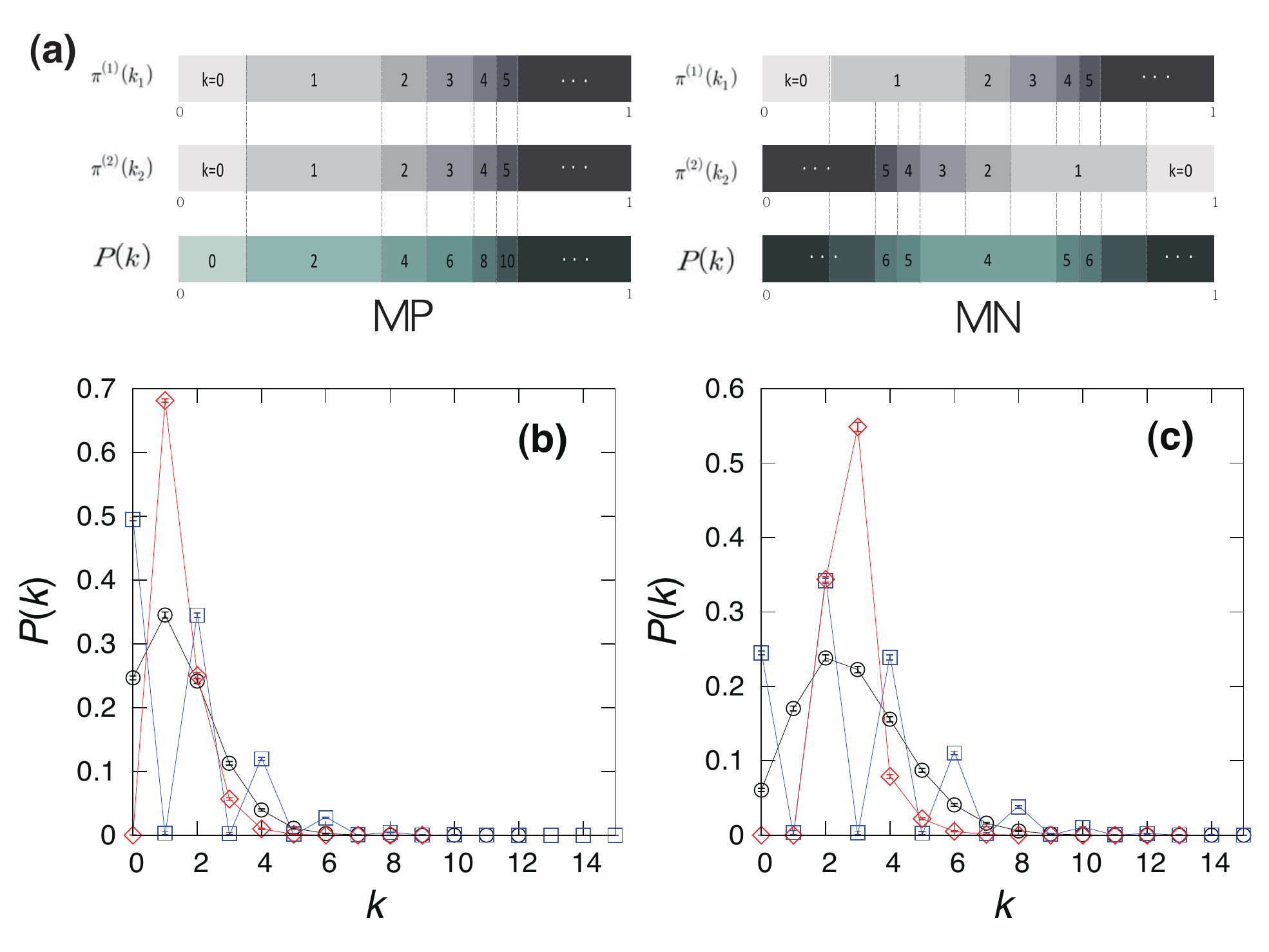}
\caption{(a) Schematic illustrations of how the mean-field-like calculation for degree distribution is done.
(b, c) Degree distributions of interlaced multiplex ER networks
with $z_1=z_2=0.7$ (b) and $z_1=z_2=1.4$ (c).
Points are numerical simulation results for the uncorrelated ($\circ$),
MP ($\Box$), and MN ($\diamond$) cases, together with 
lines representing predictions of mean-field-like calculations.
}
\end{figure}

\subsection{Uncorrelated multiplexity}
In the absence of correlation between network layers, 
the joint degree distribution factorizes, 
$\Pi_{uncorr}(k_1,k_2)=\pi^{(1)}(k_1)\pi^{(2)}(k_2)$.
The total degree distribution of the multiplex 
network is given by the convolution of $\pi^{(l)}(k_l)$,
$P_{uncorr}(k)=\sum_{k_1=0}^{k}\pi^{(1)}(k_1)\pi^{(2)}(k-k_1)$,
and its generating function $g_0^{uncorr}(x)=g_0^{(1)}(x)g_0^{(2)}(x)$,
where $g_0^{(l)}(x)$ is the generating function of $\pi^{(l)}(k_l)$.
Using $g_0(x)=e^{z(x-1)}$ for the ER network with mean degree $z$,
we have $g_0^{uncorr}(x)=e^{(z_1+z_2)(x-1)}$ for the duplex ER
network with mean intralayer degrees $z_1$ and $z_2$,
which is nothing but the generating function of an ER network
with mean degree $z_1+z_2$. Therefore, we have 
\be P(k)=\frac{e^{-z}z^k}{k!} \ee
with $z=z_1+z_2$.

\subsection{Maximally-positive correlated multiplexity}
In the maximally-positive (MP) correlated multiplex case,
a node's degrees in different layers are maximally correlated in their 
degree order; the node that is hub in one layer 
is also the hub in the other layers, and the node that has the smallest
degree in one layer also has the smallest degree in other layers.
To obtain the total degree distribution $P(k)$ of the multiplex network,
we use the following mean-field-like scheme ignoring fluctuations in the thermodynamic limit ($N\to\infty$), illustrated for the duplex
case in figure 2(a). 
We partition the unit interval into bins of sizes
$\pi^{(\ell)}(k_{\ell})$ sorted in order of $k_{\ell}$ for each $\ell$.
Combining the two partitions, we have a new partition 
which can be used to reconstruct $P(k)$ as illustrated in figure 2(a).

\subsection{Maximally-negative correlated multiplexity}
In the maximally-negative (MN) correlated multiplex case,
a node's degrees in different layers are maximally anti-correlated 
in their degree order;
a node that is hub in one layer has the smallest degree in the other layer, 
{\it etc}. The mean-field-like scheme for duplex case proceeds 
in a similar way as the MP case, except that we use two partitions
sorted in opposite orders respectively as illustrated in figure 2(a).

\begin{figure}
\hspace{1.5cm}\includegraphics[width=.7\linewidth]{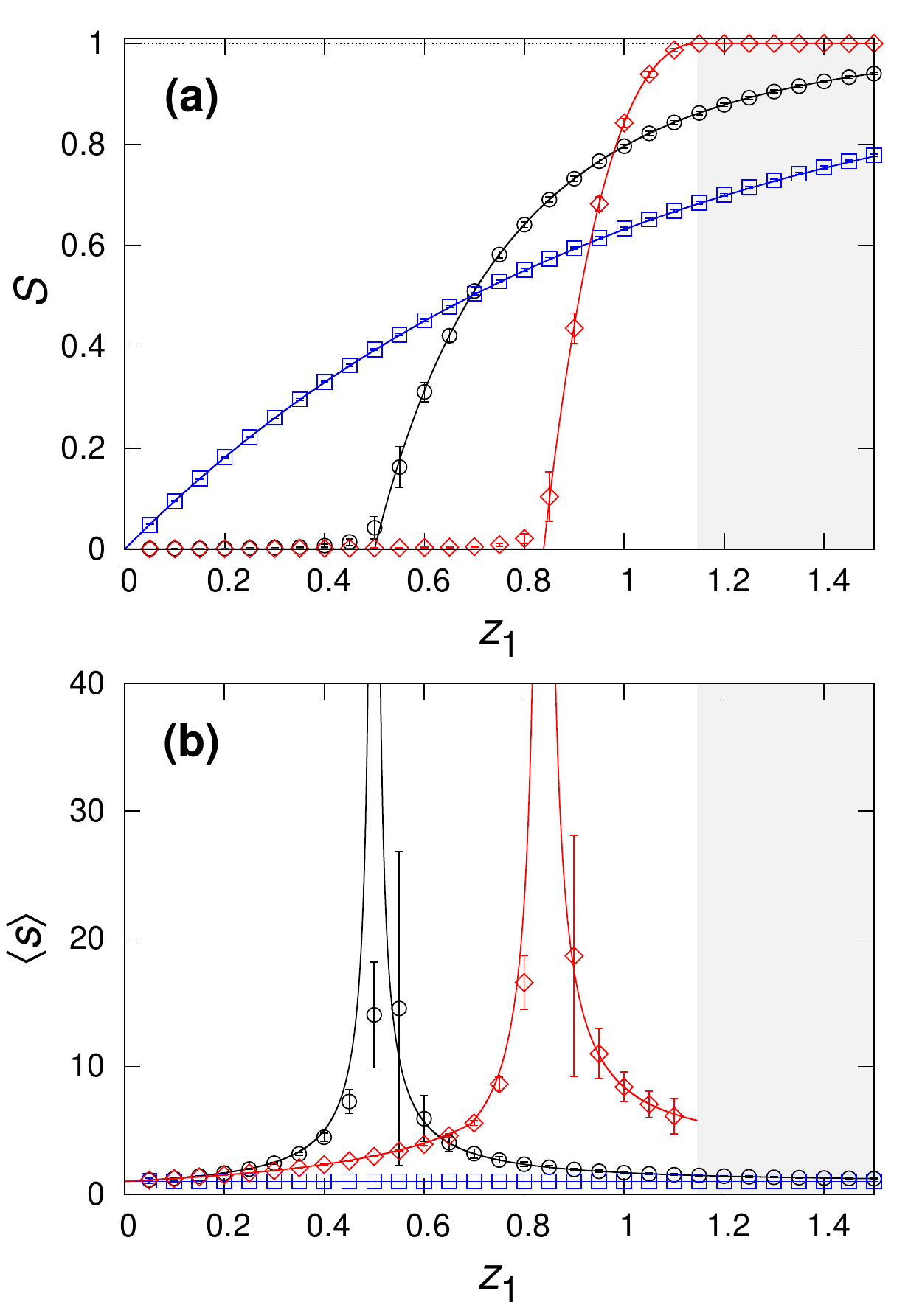}
\caption{ 
(a) The giant component size $S$  
and (b) the susceptibility $\langle s\rangle$ 
as a function of the single-layer mean degree $z_{1}$ 
of the two-layer multiplex ER networks with equal mean layer degrees for uncorrelated (black),
MP (blue), and MN (red) cases. 
Lines represent the solutions of (2) and (4) under the mean-field scheme and symbols represent the numerical simulation results obtained from 
networks of size $N=10^4$ with $10^4$ different configurations for the uncorrelated ($\circ$),
MP ($\Box$), and MN ($\diamond$) cases. Errorbars denote standard deviations.
Gray shade denotes the region in which $S=1$ for the MN case 
($z_{1}>z^*=1.14619322...$).}
\end{figure}

\section{Duplex ER networks with equal link densities}
In this section we consider ER networks with two layers of equal link densities, for which the degree distributions are most easily calculated.  

\subsection{Uncorrelated case}
The uncorrelated multiplex ER network is simply another ER network 
with mean degree $z=2z_1$, where $z_1$ is the mean degree of
the first network layer. The joint degree distribution factorizes, thus the conditional degree distribution $\Pi(k_2|k_1)$ is independent of $k_1$ and $\Pi(k_2|k_1)=\pi(k_2)$, where $\pi(k)$ denotes the Poisson distribution with mean degree $z_1$. As we increase  $z_1$,
we have a second-order percolation transition at the critical mean intralayer degree
$z_c=1/2$ \cite{newman,er}, at which the giant component emerges
in the duplex network. 
Giant component size scales in the vicinity of $z_c$ as $S\sim(z_1-z_c)^{\beta}$
with $\beta=1$, and the susceptibility 
as $\langle s\rangle \sim |z_1-z_c|^{-\gamma}$ with $\gamma=1$, 
following the standard mean-field critical behaviors \cite{newman}.

\subsection{MP case}
In this case, in the mean-field-like scheme ($N\to\infty$) each node has exactly
the same degrees in both layers, so the conditional degree distribution becomes $\Pi(k_2|k_1)=\delta_{k_2,k_1}$. Thus we have the degree distribution of the duplex network as 
\be P(k) = \left\{\begin{array}{ll} e^{-z_1}z_1^{k/2}/(k/2)! & (k~\textrm{even}), \\
0 & (k~\textrm{odd}). \end{array} \right.
\ee
Therefore, the Molloy-Reed criterion is fulfilled for all nonzero $z_1$,
as $\langle k^2\rangle -2\langle k\rangle = 4(z_1+z_1^2)-2(2z_1)=4z_1^2>0$
for $z_1\neq0$. This means that the giant component exists
for any nonzero link density, that is,
\be z_c^{MP}=0. \ee
In fact, one can obtain the solution of $S$ and $\langle s\rangle$ explicitly in this case: As $P(k)=0$ for odd $k$, (3) has $u=0$ as a nontrivial solution, 
from which it follows from (2)
\be S=1-P(0)=1-e^{-z_1},\ee which grows linearly 
with the link density near origin as $S \sim z_1$, that is $\beta=1$.
From $u=0$, the susceptibility $\langle s\rangle=1$ for
all $z_1>0$. 
This means that only isolated nodes are outside the giant component
and all the linked nodes form a single giant component.
All these predictions are confirmed by numerical simulations (figure 3).

\subsection{MN case}
In this case one can easily show that distinct regimes appear as $z_1$
increases. Among them, three regimes are of relevance for the giant component properties: {\it i)}
For $0\le z_1\le\ln2$, more than half of nodes are of degree zero in each layer so every linked node in one layer is coupled with a degree-$0$ node in the other layer under MN coupling. In this regime the conditional degree distribution takes a rather complicated form 
\be
\Pi(k_2|k_1)=\left\{\begin{array}{ll}
\left[2\pi(0)-1\right]/\pi(0) & (k_2=0,k_1=0),\\
\pi(k_2)/\pi(0) & (k_2\neq0,k_1=0),\\
\delta_{k_2,0} & (k_1\neq0),
\end{array}\right.
\ee 
and thus 
$P(k)$ is given by 
\be
P(k)=\left\{\begin{array}{ll}
2\pi(0)-1 & (k=0),\\
2\pi(k) & (k\ge1).
\end{array}\right.
\ee
In this regime there is
no giant component. 
{\it ii)} For $\ln2\le z_1\le z^*$, similar consideration leads to $\Pi(k_2|k_1)$ and $P(k)$ given by
\numparts
\begin{equation}
\Pi(k_2|0)=\left\{\begin{array}{ll}
0 & (k_2=0),\\
\left[2\pi(0)+\pi(1)-1\right]/\pi(0) & (k_2=1),\\
\pi(k_2)/\pi(0) & (k_2\geq2),
\end{array}\right.
\end{equation}
\begin{equation}
\Pi(k_2|1)=\left\{\begin{array}{ll}
\left[2\pi(0)+\pi(1)-1\right]/\pi(1) & (k_2=0),\\
\left[1-2\pi(0)\right]/\pi(1) & (k_2=1),\\
0 & (k_2\geq2),
\end{array}\right.
\end{equation}
\begin{equation}
\Pi(k_2|k_1)=\delta_{k_2,0} \qquad\qquad\qquad\qquad (k_1\geq2),
\end{equation}\endnumparts
and  
\be P(k)=\left\{\begin{array}{ll}
0 & (k=0), \\
2[2\pi(0)+\pi(1)-1] & (k=1),\\
2\pi(2)-2\pi(0)+1 & (k=2), \\
2\pi(k) & (k\ge3).
\end{array}\right.\ee
In this regime, $\langle k^2\rangle-2\langle k\rangle=
2(z_1^2-z_1-2e^{-z_1}+1)$, which becomes positive for $z_1>z_c^{MN}$
where
\be z_c^{MN}=0.838587497... \ee
Therefore the giant component emerges at a significantly higher
link density than the uncorrelated multiplex case.
Being delayed in its birth, the giant component grows more abruptly
once formed [see figure 3(a)] than the other two cases.
This regime is terminated at $z=z^*$, determined by the condition 
$2\pi(0)+\pi(1)=1$, from which we have $z^*=1.14619322...$ 
{\it iii)} For $z_1\ge z^*$, we have $P(0)=P(1)=0$. In that case,
we have $u=0$ from (3) and thereby $S=1$ from (2).
This means that the entire network becomes connected into a single component
at this finite link density, which can never be achieved for 
ordinary ER networks.
Despite these abnormal behaviors and apparent differences in the steepness of the order parameter curve near $z_c$, the critical behavior is found to be consistent with that of standard mean-field, $\beta=1$ and $\gamma=1$ (figure 4).

\begin{figure}
\flushright\includegraphics[width=.95\linewidth]{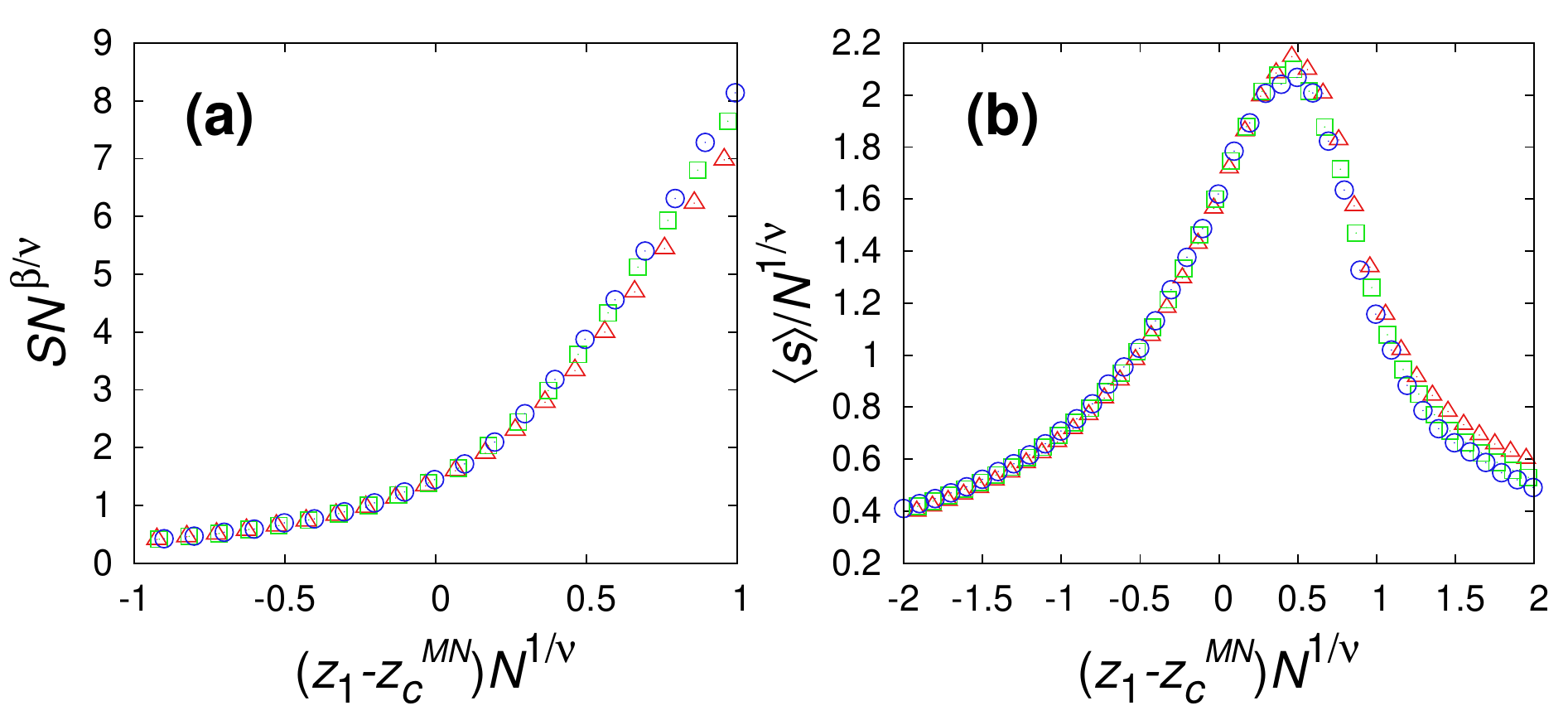}
\caption{
Data collapse of (a) scaled giant component sizes, $SN^{\beta/\nu}$, and (b) scaled susceptibility, $\langle s\rangle/N^{1/\nu}$, vs.\ $(z-z_c^{MN})N^{1/\nu}$ for the MN case, with $\beta=1$ and $\nu=3$, consistent with the conventional mean-field behaivor $\beta=1$ and $\gamma=1$. $N$ is $10^4 (\triangle)$, $10^5 (\Box)$, and $10^6 (\circ)$. 
}
\end{figure}

\section{Imperfect correlated multiplexity}

In the previous section we have seen that {\it maximally} correlated
or anti-correlated multiplexity affects the onset of emergence 
of giant component in multiplex ER networks.
Despite its mathematical simplicity and tractability,
in real-world multiplex systems the correlated multiplexity 
would hardly be maximal. Therefore it is informative to see
how the results obtained for the maximally
correlated multiplexity are interpolated when the system possesses
partially correlated multiplexity.

\begin{figure}
\hspace{1.5cm}\includegraphics[width=.8\linewidth]{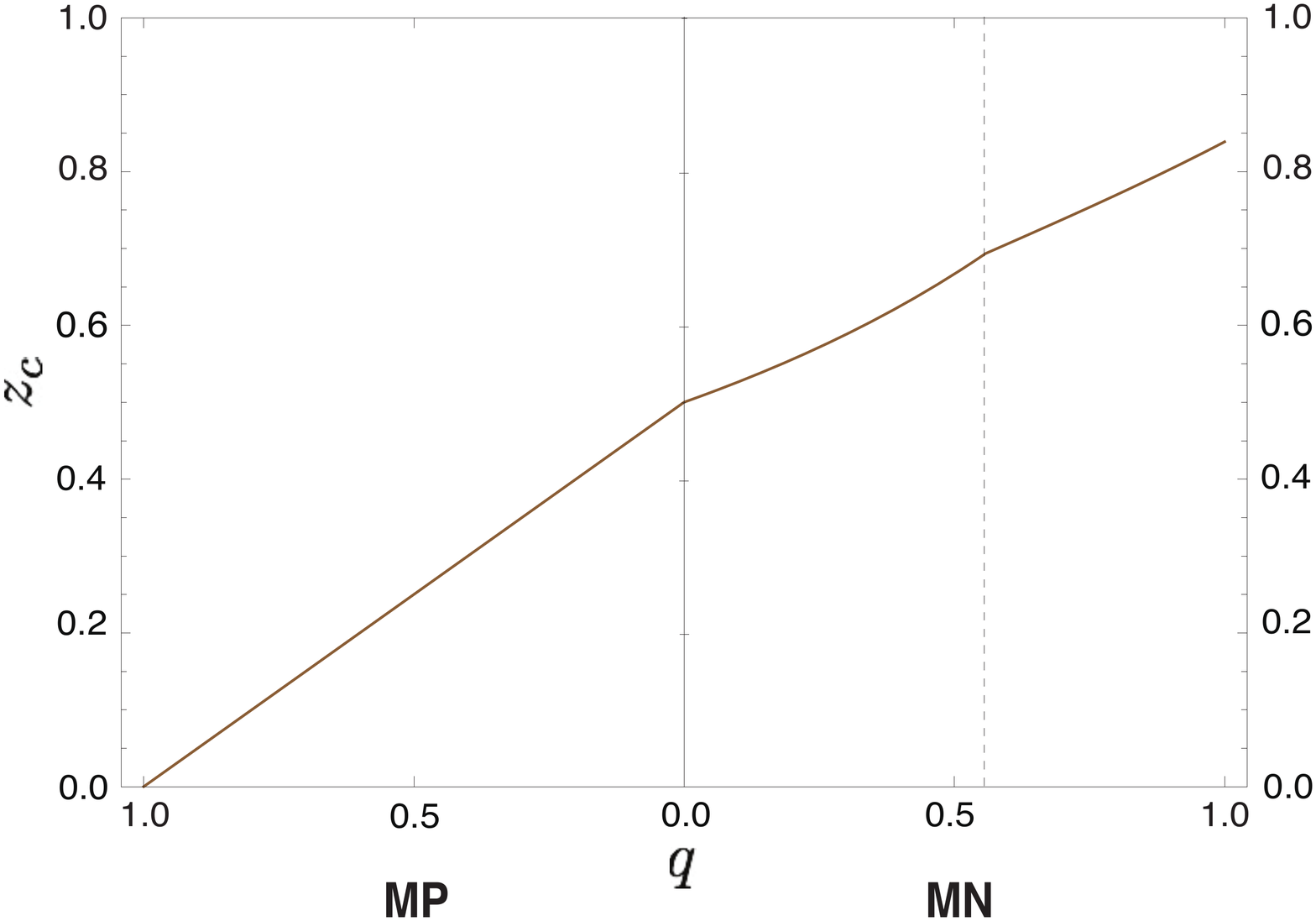}
\caption{
Plot of equation (15) for the critical mean degree $z_{c}$ as a function of $q$, the fraction of correlated multipex nodes. The cases $q=1$ denote maximally correlated multiplexity, 
$0<q<1$ partially correlated multiplexity, 
and $q=0$ uncorrelated multiplextiy.
The vertical dotted line is drawn at $q=2-1/\ln2$ across which $z_c$ takes different formulae in equation (15b).
}
\end{figure}

To this end, we consider duplex ER networks where a fraction $q$ of nodes 
are maximally correlated multiplex while the rest fraction $1-q$ 
are randomly multiplex.
Then the degree distribution of the interlaced network is given by
$P_{partial}(k)=qP_{maximal}(k)+(1-q)P_{uncorr}(k)$, where $maximal$ is
either $MP$ or $MN$.
Following similar steps as in the previous section we obtain the critical
link density as a function of $q$ as 

\numparts
\be
z_c = (1-q)/2 
\ee
for positively correlated case and
\be
z_c = \left\{ \begin{array}{ll}
1/(2-q) & \textrm{\qquad $(q<2-1/\ln2)$},\\
z_{1}(q) & \textrm{\qquad $(q>2-1/\ln2)$}
\end{array} \right.
\ee
\endnumparts 
for negatively correlated case, where $z_{1}(q)$ is the solution of $(2-q)z_{1}^{2}-z_{1}-2qe^{-z_{1}}+q=0$.
In figure 5, we show the plot of $z_c$ as a function of $q$ given by (15).
This result shows that the effect of correlated multiplexity is not only
present for maximally correlated cases but for general $q$.

\section{Duplex ER networks with general link densities}

In the previous sections we focused on the cases with $z_1=z_2$.
In this section we consider general duplex ER networks with $z_1\neq z_2$.
We performed numerical simulations with $N=10^4$ for MP, uncorrelated,
and MN correlated multiplex cases with $z_1$ and $z_2$ in the range from $0$ to $3$.
In figure 6, the giant component size $S$ for general $z_1$ and $z_2$
are shown. Similarly to the $z_1=z_2$ cases, the giant component emerges
at lower link densities for the MP case but grows more slowly than
the uncorrelated case, whereas it emerges at higher link densities
for the MN case but grows more abruptly and connects all the nodes
in the network at finite link density. Therefore the effect of
correlated multiplexity is qualitatively the same and generic.

\begin{figure}
\includegraphics[width=\linewidth]{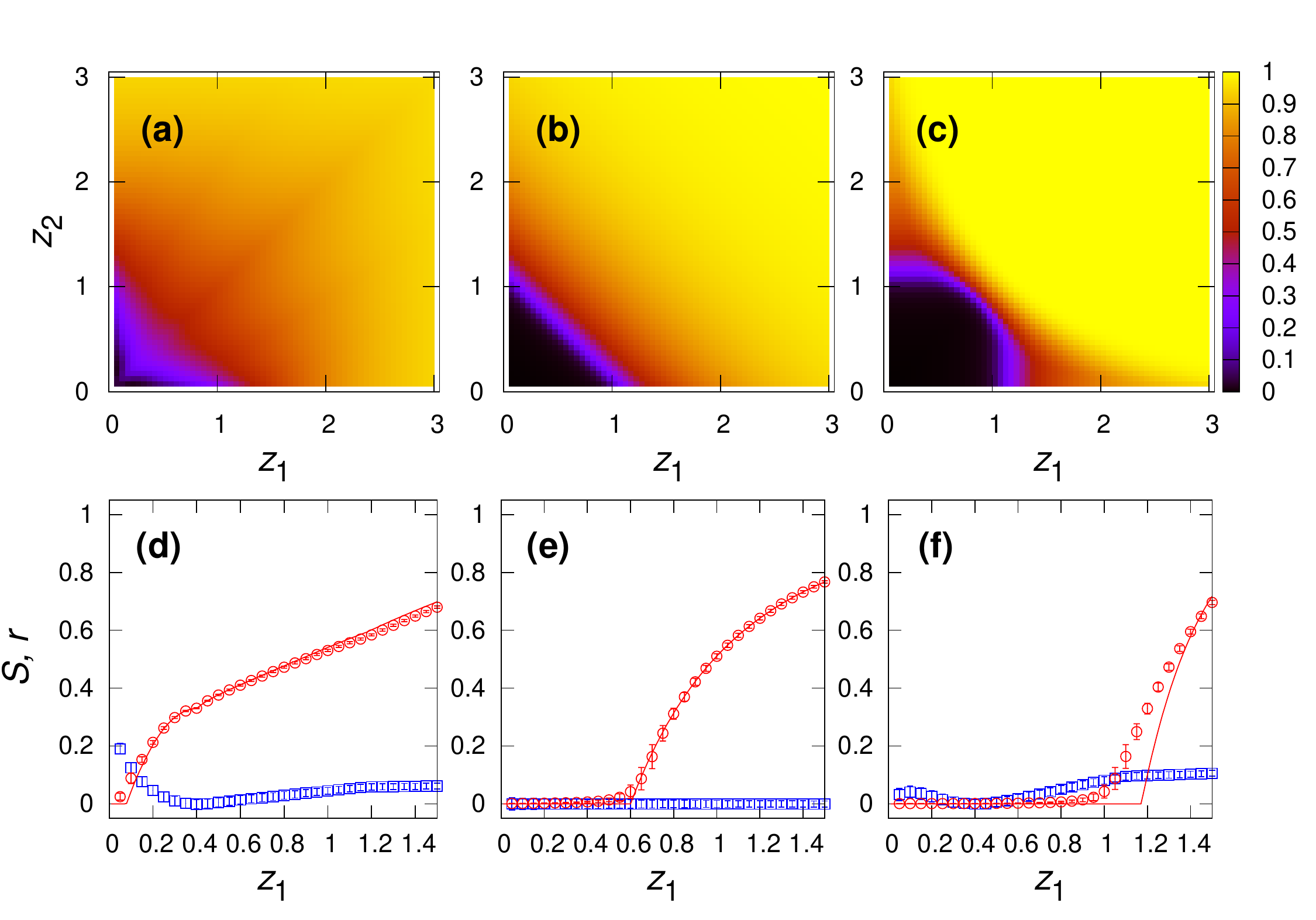}
\caption{Numerical simulation results of 
the size of giant component of duplex ER networks
with (a) MP, (b) uncorrelated, and (c) MN cases.
In (d--f), the giant component size $S$ (red) is plotted for $z_2=0.4$,
along with the assortativity coefficient $r$ (blue) for the MP (d), 
uncorrelated (e), and MN (f) cases. Mean-field-like calculation
results (lines) deviates from the numerical simulation results ($\circ$)
when assortativity coefficient ($\Box$) becomes nonzero, that is,
the network display degree-degree correlation.
Errorbars denote standard deviations from $10^4$ independent runs.
}
\end{figure}

It is worthwhile to note, however, that although
the mean-field-like approximation for $P(k)$ introduced in section 3,
which was very effective for equal link densities, still yields a qualitatively correct picture, it 
fails in quantitative agreement with numerical simulation results
for general $z_1\neq z_2$.
To understand the origin of this discrepancy, we consider network correlations.
Assortativity coefficient $r$ \cite{assort} defined as 
the Pearson correlation coefficient between
the (total) degrees of nodes connected by a link in the network measures
the degree-degree correlations in the network at the two-node level.
Nonzero assortativity is known to alter 
the connectivity properties of networks \cite{assort}.
To see the role of degree-degree correlations induced by
correlated multiplexity, in figure 6(d--f) we plot the giant component
size obtained from numerical simulations and mean-field calculations,
together with the assortativity coefficient as a function $z_1$ with 
$z_2=0.4$ fixed.
With uncorrelated multiplexity, the degree-degree correlation is absent,
$r=0$, and the numerical simulation and mean-field calculation agree
perfectly [figure 6(e)].
For MP and MN cases, however, the assortativity of multiplex network is generically nonzero (positive), that is the multiplex network becomes correlated, which is responsible for the deviations between the numerical simulation and mean-field calculation results. This discrepancy vanishes at $z_1=z_2$, at which the assortativity also vanishes [figures 6(d,f)].
Note that there was no degree-degree correlations at the individual
network layer level. 
Therefore this result shows that generically correlated multiplexity not only
modulate $P(k)$ but also introduce higher-order correlations in 
the multiplex network structure. 
Meanwhile, it is worthwhile to note that such a multiplexity-induced degree correlation has similar origin as the correlation in colored-edge networks recently studied in a different context of network clustering \cite{gleeson}.

\section{Conclusion}

In conclusion, we have studied the effect of correlated multiplexity 
on the structural properties of multiplex network system,
a better representation of most real-world complex systems
than the single, or simplex, network. 
We have demonstrated 
that the correlated multiplexity
can dramatically change the giant component properties.
With positively correlated multiplexity, the giant component 
emerges at a much lower critical link density, which even approaches to zero 
for MP case, than for uncorrelated multiplex cases.
Once formed, however, the giant component grows much more gradually.
With negatively correlated multiplexity, the giant component
emerges at a much higher critical density than for uncorrelated multiplex
cases, but once formed it grows
more abruptly and can establish the full connectivity to connect the entire network into a single component
at a finite link density.
These results show that a multiplex complex system
can exhibit structural properties that cannot be represented by
its individual network layer's properties alone, the impact of which on network dynamics is to be explored in future study.

\ack
We are grateful to J. P. Gleeson for useful comments and informing us of Ref.~\cite{gleeson}. This work was supported by Mid-career Researcher Program
(No.~2009-0080801) and Basic Science Research Program (No.~2011-0014191)
through NRF grants funded by the MEST.
K-ML is also supported by Global PhD Fellowship Program (No.~2011-0007174) through NRF, MEST.

\end{document}